\documentstyle[editedbook,epsfig,psfig,epsf]{mq}
\def\eg{{\it e.g.} }
\def\etal{{\em et al.} }
\def\ie{{\em i.e.} }

\def\lsim{\lower.5ex\hbox{$\; \buildrel < \over \sim \;$}}
\def\gsim{\lower.5ex\hbox{$\; \buildrel > \over \sim \;$}}

\def\cm2{cm$^2$ }
\def\se1{s$^{-1}$ }

\begin{opening}
\title{Radiatively Driven Jets around Black Holes}
\author{Indranil Chattopadhyay$^1$ \& Sandip. K. Chakrabarti$^{1,2}$}
\institute{$^1$ S. N. Bose National Centre for Basic Sciences, Block-JD, Sector-III, Kolkata-98 India. \\
$^2$ Also at Centre for Space Physics, P-61 Southend Gardens, Kolkata-84, India. }
\end{opening}

\runningtitle{Radiatively Driven Jets}
\runningauthor{Chattopadhyay \& Chakrabarti}
\begin{document}
\vspace{-0.5cm}
\begin{abstract}
{\small The hot, puffed
up, post-shock region of an advective disc is the
source of high energy photons and also the jets and outflows. 
We study the relativistic equations of motion of 
jets as these high energy photons interact with
them. We show that the much discussed terminal velocity of jets depends
on the comparative value of radiative energy density, flux and the radiative 
pressure. We show that electron-positron pair plasma jets achieves highly
relativistic terminal speeds for higher disc luminosities.} 
\end{abstract}

\section{Introduction}

Rotating matter while falling onto black holes creates a temporary depository
of matter called {\it accretion disc}. As matter moves closer to the 
black hole, 
at around a few tens of Schwarzschild radii ($r_g$), centrifugal force
tends to be comparable to the inward gravitational force and the supersonic
inflow is slowed down. If this slowing down occurs in a thin region 
\cite{C89}, the flow is said to suffer a {\it shock}, where the Mach number 
jumps discontinuosly from supersonic to subsonic branch. 
This slowed down post-shock flow is considerably heated up
and as a result, puffs up (to maintain hydrostatic balance along
z-direction) in the form of a tori called CENBOL (centrifugal pressure 
supported boundary layer). CENBOL contains a copius hot 
electrons which inverse-Comptonize the soft photons from outer cool thin disc
\cite{SS73}, producing the hard-power law tail \cite{CT95}. As a lot of heat 
is stored in the CENBOL, Chakrabarti and his
co-workers (\eg \cite{C99}, \cite{DC99}, \cite{DCNC01}) 
has shown that the thermal pressure along the
vertical direction pushes the matter out in the form of jets. 
In this paper, we investigate the {\it interaction} of these {\it jets} with 
the high energy {\it photons} {\it radiated} by the CENBOL.

Interaction of radiation and jets have been studied by several workers.
Piran (\cite{P82}) showed that it is difficult to accelerate outflows from
thick accretion disc beyond Lorentz factor ${\gamma}>1.5$.  
Icke (\cite{I89}) in a very important paper
showed that, if outflows are to be accelerated by radiation from `infinite'
thin discs, the terminal speed (`magic speed' in his parlance)
achieved is just $0.451c$, where $c$ is the velocity of light. Fukue 
(\cite{F96}), showed the terminal speed for rotating flow above thin disc,
is little less than 
what Icke had obtained. The conclusion is that
radiative interaction actually limits the jet terminal speed. 
We improve these works by studying the interaction of a
jet with radiation coming from the most general form of accretion flow, namely
the {\it advective discs} (\eg \cite{C89},\cite{CT95},\cite{C96b}).
The first three moments of the radiation intensity for 
\eg, the radiation energy density,
radiative flux and various components of radiative pressure are being
calculated following the treatment of Chattopadhyay \& Chakrabarti 
(\eg \cite{CC00}, \cite{CC02}). For simplicity, we assume that
the radiation originates only from the CENBOL. 

In the next section we show that, the terminal speed achieved
by the outflowing plasma depends on the comparative value of the various
moments of radiation involved. We show that the radiation from
the advective flows can produce relativistic 
jets. In \S3, we draw our conclusions.

\begin{figure}[htb]
\vbox{
\centerline{
\psfig{file=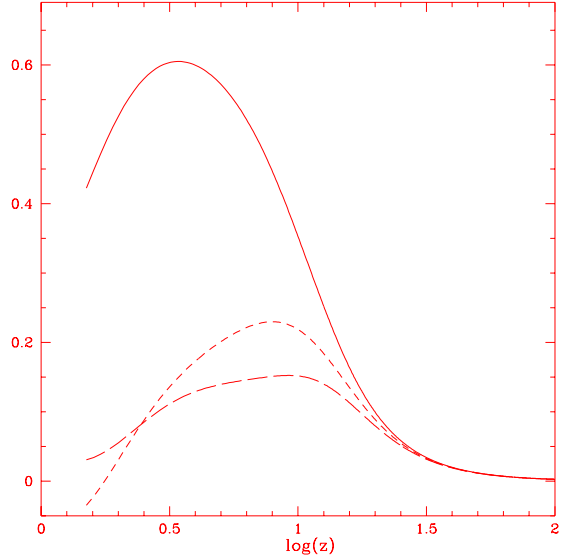,width=8cm}}
\caption{Comparison of radiation field quantities ${\cal E}$ (solid), 
${\cal F}$ (dashed),
and ${\cal P}$ (long-dashed) plotted with $log(z)$ in non-dimensional
units. These field quantities are calculated for disc luminosity $L=0.06
L_{Edd}$. Radial dimension of CENBOL (i.e., shock location) is
chosen to be $10r_g$.}}
\label{fig:ex}
\end{figure}

\section{Equations of motion and radiative acceleration of jets}

We assume that the electron-positron plasma 
jets are non-rotating and confined along the
axis of symmetry. To study only the radiation effects, we ignore the
gas pressure gradient term. We confine our investigation within
the realm of special relativity and the effects of strong
gravity is taken care by Paczy\'nski-Wiita potential (\eg \cite{PW80}).
The unit of length, time, mass
and velocity are $r_g=2GM_B/c^2$, $2GM_B/c^3$, $M_B$ and $c$
respectively, where $G$ and $M_B$ are the gravitational constant and mass of
the central black hole. The metric considered is given by,
$d{\tau}^2=dt^2-dr^2-r^2d{\phi}^2-dz^2.$
We assume $u^r=u^{\phi}=0$ and also ${\partial}/{\partial}t=
{\partial}/{\partial}r={\partial}/{\partial}{\phi}=0$. We
define a 3-velocity $v$, given by $v^2=-(u_iu^i)/(u_tu^t)$, where
the latin index (\eg $i,j,k$) denotes space variables.
The Lorentz factor is given by, ${\gamma}^2=1/(1-v^2)$. Hence in our case 
$u^z={\gamma}v$, $u_z=-{\gamma}v$, $u^t=u_t={\gamma}$.
The equation of motion (see, Mihalas and Mihalas \cite{MM84} and
Kato \etal \cite{KFM98}) is given by,
$$
\frac{du^z}{d{\tau}}=-\frac{1}{2(z-1)^2}+{\kappa}_{es}[{\gamma}F^z
-{\gamma}^2Eu^z-P^{zz}u^z+u^z(2{\gamma}F^zu^z-P^{zz}u^zu^z)],
\eqno{(1)}
$$
where, ${\kappa}_{es}$, $F^z$, $E$, $P^{zz}$ are the Thomson scattering 
opacity, the z-component of radiative flux on the
z-axis, energy density on the axis of symmetry, and the z-z component of
pressure tensor on the axis of symmetry. The term $-1/\{2(z-1)^2\}$ is the
gravity term with Paczy\'nski-Wiita potential. 
Using the expressions of $u^z$ and $u^t$
we can reduce Eq. (1) in terms of the 3-velocity $v$, and is given by,
$$
\frac{dv}{dz}=\frac{-\frac{1}{2(z-1)^2}+[{\gamma}{\cal F}-{\gamma}^3v{\cal E}
-{\gamma}v{\cal P}+{\gamma}^3(2v^2{\cal F}-v^3{\cal P})]}
{{\gamma}^4v},
\eqno{(2)}
$$
where, ${\cal F}$, ${\cal E}$, and ${\cal P}$ are radiative flux,
energy density and z-z component of pressure tensor which has been multiplied
by ${\kappa}_{es}$.
To find the expression for terminal speed, we put the term in the square 
bracket of Eq. (2) equal to zero and get a quadratic equation for $v_t$,
$$
{\cal F}v^2_t-({\cal E}+{\cal P})v_t+{\cal F}=0.
$$
It is easy to show that at $z{\rightarrow}$large, 
${\cal P}{\lsim}{\cal F}{\lsim}{\cal E}$, and also that all three
quantities vary slowly with $z$ (Fig. 1). 
Thus for large $z$, let ${\cal P}=A$, hence
${\cal F}=A+{\delta}$ and ${\cal E}=A+{\eta}$, where ${\delta}{\ll}A$
and ${\eta}{\ll}A$ with ${\delta}{\lsim}{\eta}$.
Hence, from the above quadratic equation we get,
$$
v_t=\frac{({\cal E}+{\cal P})-{\sqrt{({\cal E}
+{\cal P})^2-4{\cal F}^2}}}{2{\cal F}}
\eqno{(3)}
$$  
Putting the radiation field quantities in terms of $A$ in Eq. (4), 
we have,
$$
v_t{\approx}\frac{1+{\eta}/(2A)}{1+{\delta}/A}{\lsim}1
\eqno{(4)}
$$
Equation (4) shows that $v_t$ can be very close to the velocity of light, thus
radiation drag does not limit the terminal velocity to some moderately
relativistic values. In case of gas
flows above a thin disc (\eg \cite{I89},\cite{F96}), ${\cal P}=1/3
{\cal E}$ and ${\cal E}{\sim}2{\cal F}$, putting those values in Eq. (4),
we have $v_t=\frac{1}{3}(4-{\sqrt{7}})$ (see, Icke\cite{I89}). 
{\it We have thus proved that the terminal velocity achieved
depends on the radiation field produced by the structure of the disc}.
If we now integrate Eq. (2) with a very small injection velocity, we can find
the solution topologies for jets as shown in Fig. 2a.
Though there is no upper limit to $v_t$, even then increasing the luminosity,
will not increase $v_t$ drastically, as there is ${\gamma}^4$
term in the RHS of the denominator of Eq. (2), and hence as $v{\rightarrow}1$,
$dv/dz{\rightarrow}0$. In Fig. 2b the velocity at $z=10000r_g$ \ie, 
$v_{10000}$ is plotted with the disc luminosity $L$ and is found that as
one increases $L$, the response of $v_{10000}$ is not linear. Figure 2b shows
that it is difficult to get terminal speeds ${\sim}0.99c$ with radiative
acceleration only, as that would require extremely high luminosity, which
is difficult to produce without, at the same time, cooling the
CENBOL itself.

\begin{figure}[htb]
\vbox{
\centerline{
\psfig{file=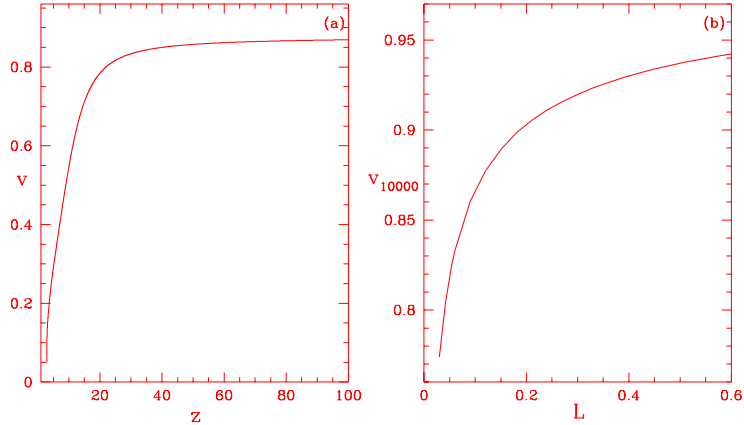,width=10cm}}
\vspace{-4cm}
\caption{(a) Velocity profile of an electron-positron jet, with
input parameters $z_{in}=3r_g$, $v_{in}=0.05c$. The radiation field
quantities correspond to disc luminosity $0.06L_{Edd}$. (b) The variation of
$v_{10000}$ (\ie $v$ at $z=10000r_g$) with the accretion disc luminosity $L$ 
in units of $L_{Edd}$.}}
\label{fig:ex}
\end{figure}

\section{Discussion and Concluding Remarks}

In our investigation we have assumed the entire disc luminosity is coming
only from the CENBOL region.
It can be shown elsewhere that the contribution from the pre-CENBOL
thin disc will marginally affect the results shown here. 
We conclude that, the terminal speeds calculated previously, by using
radiation fields from very specialised accretion discs (\eg, thin disc,
thick disc, etc), show mildly relativistic flows, and are basically the
artefacts of those disc geometries themselves. If one considers
more generic disc model as we have done, the
radiation drag does not introduce a mildly
relativistic upper limit for terminal speed. We find that achieving terminal
velocity close to that of light by radiation pressure effects is possible.

\end{document}